\documentclass[12pt]{article}
\usepackage{graphicx}
\usepackage{epstopdf}
\title{Boundary chromatic polynomial}

\author{Jesper Lykke Jacobsen${}^{1,2}$ and
        Hubert Saleur${}^{2,3}$ \\[2.0mm]
$^{1}$Universit\'e Pierre et Marie Curie, 4 place Jussieu, \\
75252 Paris Cedex 05, France \\
$^{2}$ Institut de Physique Th\'eorique, CEA Saclay, \\
91191 Gif-sur-Yvette, France \\
${}^3$ Department of Physics and Astronomy, \\
University of Southern California, \\
Los Angeles, CA 90089, USA}

\begin{document}

\maketitle

\begin{abstract}

  We consider proper colorings of planar graphs embedded in the
  annulus, such that vertices on one rim can take $Q_{\rm s}$ colors,
  while all remaining vertices can take $Q$ colors. The corresponding
  chromatic polynomial is related to the partition function of a
  boundary loop model. Using results for the latter, the phase diagram
  of the coloring problem (with real $Q$ and $Q_{\rm s}$) is inferred,
  in the limits of two-dimensional or quasi one-dimensional infinite
  graphs.  We find in particular that the special role played by
  Beraha numbers $Q=4\cos^2{\pi\over n}$ for the usual chromatic
  polynomial does not extend to the case $Q\neq Q_{\rm s}$.  The
  agreement with (scarce) existing numerical results is perfect;
  further numerical checks are presented here.

\end{abstract}

\section{Introduction}

Let $G=(V,E)$ be a planar graph embedded in the annulus. Let $V_{\rm
  s} \subseteq V$ be the subset of vertices surrounding the face that
contains the point at infinity. In other words, $V_{\rm s}$ are the
vertices on the outer rim of the annulus. Place a spin variable
$\sigma_i=1,2,\ldots,Q$ on each bulk vertex ($i \in V \setminus V_{\rm
  s}$) and a boundary spin $\sigma_j=1,2,\ldots,Q_{\rm s}$ on each
boundary vertex ($j \in V_{\rm s}$). We suppose initially that $Q_{\rm
  s} \le Q$, so that $Q-Q_{\rm s}$ of the colors allowed for the bulk
spins are forbidden for the boundary spins.

The Potts model partition function $Z_G(Q,Q_{\rm s};{\bf v})$---also
known to graph theorists as the multivariate Tutte polynomial---can be
defined through a slight generalization of the usual Fortuin-Kasteleyn
expansion \cite{Kasteleyn}
\begin{equation}
 Z_G(Q,Q_{\rm s};{\bf v}) = \sum_{A \subseteq E} Q^{k(A)}
 \left( \frac{Q_s}{Q} \right)^{k_{\rm s}(A)} \prod_{e \in A} v_e
 \label{FK}
\end{equation}
where $k(A)$ is the number of all connected components (clusters) in
the graph induced by the edge subset $A$, and $k_{\rm s}(A)$ is the
number of connected components that contain at least one vertex from
$V_{\rm s}$.  In other words, $Q_{\rm s}$ (resp.\ $Q$) is the weight
of a cluster that contains at least one (resp.\ does not contain any)
vertex in $V_{\rm s}$. The edge variables ${\bf v} = \{v_e\}_{e \in
  E}$ are related to the usual spin-spin couplings $K_e$ through the
relation $v_e = \exp(K_e)-1$.

Note that in (\ref{FK}) there is no need for $Q$ and $Q_{\rm s}$ to be
integers, nor do we have to impose the constraint $Q_{\rm s} \le Q$.
We shall henceforth promote (\ref{FK}) to the {\em definition} of the
(boundary) Potts model \cite{JS1}.

In this paper we wish to study the problem of proper colorings of $G$,
such that bulk vertices can have $Q$ different colors, whereas boundary
vertices can have only a subset of $Q_{\rm s}$ colors. Adjacent
vertices (of whatever type) are constrained to have different colors.
The partition function $Z_G(Q,Q_{\rm s};-1)$, i.e. with all $v_{\rm e}=-1$,
counting the number of such proper colorings is referred to as the
{\em boundary chromatic polynomial} and denoted $P_G(Q,Q_{\rm s})$.
Note that $P_G(Q,Q)$ is nothing else than the usual
chromatic polynomial, which has been studied extensively in
the literature \cite{Sokal1}.

We address in particular the issue of the phase diagram of
$P_G(Q,Q_{\rm s})$ for ``large graphs''---what is meant precisely by
this will be discussed below. The main result is the location and
nature of a series of phase transition (with corresponding behaviors
of zeroes of $P_G$) occurring when one varies one or both of the
parameters $Q$ and $Q_{\rm s}$. We emphasize that most of our results
are quite general, and do not depend on the detailed structure of the
underlying graph $G$.

In the usual case $Q=Q_{\rm s}$, one of the striking features of the
chromatic polynomial is that for ``large graphs'' its real zeroes
possess accumulation points which belong to the magic set of Beraha
numbers:
\begin{equation}
 B_t = 4 \cos^2 \left(\frac{\pi}{t} \right) \quad
 \mbox{for } t=2,3,\ldots \,.
 \label{Berahano}
\end{equation}
Note that the first two such numbers are $Q=0$ and $Q=1$, which are
usually exact zeroes for finite graphs as well. One of the striking
conclusions of our study is that the special role played by Beraha
numbers is not very resistant to changing $Q_{\rm s}$. Depending on
the problem one choses, there can indeed be accumulating zeroes at
other special points of the real axis.

\begin{figure}
\centering
\includegraphics[scale=.4] {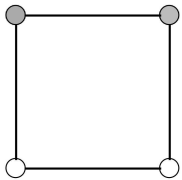}
\caption{For this graph (with the shaded vertices on the boundary) the
  chromatic polynomial continued from the region $Q\geq Q_{\rm s}$ is
  $P_G=(Q^2-3Q+3)Q_{\rm s}(Q_{\rm s}-1)$. It does not vanish for $Q=0$
  nor $Q=1$. Meanwhile the chromatic polynomial continued from the
  region $Q\leq Q_{\rm s}$ is obtained---in this simple example---by
  exchanging $Q$ and $Q_s$ in the above expression, and does vanish
  for $Q=0$ and $Q=1$.}
\label{fig:square}
\end{figure}

It is important to realize that the definition of $P_G(Q,Q_{\rm s})$,
albeit very natural, can lead to counterintuitive features in
particular when interpreted outside the initial domain of definition
$Q_{\rm s}\leq Q$. For instance it turns out that for most graphs,
$P_G(Q,Q_{\rm s})$ does not vanish when $Q=0$ or $Q=1$, even though in
that case there is no way---forgetting the boundary contribution---to
color the bulk vertices with $Q$ colors. The point is that in the
original definition (\ref{FK}), spins belonging to clusters that touch
the boundary receive a fugacity $Q_{\rm s}$, which initially is
assumed smaller or equal to $Q$, but which, after continuation, can in
fact be greater, hence ``pumping'' the number of colors in the bulk.
Fig.~\ref{fig:square} provides a simple example of this subtlety.

One could define another chromatic polynomial starting from the
situation where $Q\leq Q_{\rm s}$. In terms of the subsequent cluster
and loop model expansions, it would however be much less
interesting. Indeed, in such a model, only clusters not containing any
of the bulk spins would get the fugacity $Q_{\rm s}$, and thus only
loops ``glued to the boundary'' would get a fugacity different from
the bulk ones. This presumably would not affect the patterns of
zeroes.

The layout of the paper is as follows. In section \ref{sec:BLM} we
relate the boundary chromatic polynomial to a loop model which was
previously introduced in \cite{JS1} and further studied in
\cite{JS2}. In section \ref{sec:BKW} the issue of the phase diagram  is
transposed into the setting of the Beraha-Kahane Weiss theorem
\cite{BKW} which we review.  The necessary input for applying that
theorem is supplied by an analytic continuation of the field theoretic
results of \cite{JS1}, as explained in section
\ref{sec:infiniteW}. Here we also arrive at the main results of the
paper, which are the phase transition loci
(\ref{h_criterion})--(\ref{D_criterion}). All of this applies to the
two-dimensional thermodynamic limit. However, the main results remain
valid for quasi one-dimensional graphs, and we provide the necessary
arguments in section \ref{sec:finiteW}. A few numerical validations of
our results are given in section \ref{sec:numerics} after which we
present our conclusions.

\section{Boundary loop model}
\label{sec:BLM}

The cluster model (\ref{FK}) can obviously be defined for any graph $G$.
However, when $G$ is planar, the cluster model can be turned into a
loop model on the medial graph ${\cal M}$.

We recall that the medial (or surrounding) graph has a vertex
standing on each edge $e \in E$, and an edge between vertices standing
on edges $e_1,e_2$, whenever $e_1,e_2$ are incident to a common vertex
in $V$ and surround a common face in $G$.

\begin{figure}
 \centering
 \includegraphics[width=130pt]{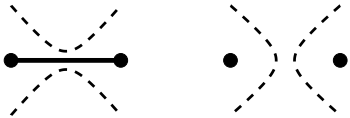}
 \caption{The transition system (shown as dashed lines) depends on
   whether a given edge $e$ (show as a solid line) is present in [left
   panel] or absent from [right panel] the edge subset $A \subseteq E$
   in (\ref{BLM}).}
 \label{fig:trans}
\end{figure}

A non-intersecting transition system on ${\cal M}$ is defined locally
as in Fig.~\ref{fig:trans}. Globally, this transition system is a set
of loops---or cycles in the standard graph theoretical
terminology---which separate clusters in $G$ from their duals. By the
existence of a point at infinity, the inside and outside of a loop are
well defined. A loop that contains at least one vertex of $V_{\rm s}$
on its inside is called a boundary loop. A loop that is not a boundary
loop is called a bulk loop.

Let now $\ell(A)$ be the total number of loops, and $\ell_{\rm s}(A)$ the
number of boundary loops. By the Euler relation, one has
$k(A) = \frac12(\ell(A) + |V| - |E|)$, so that
\begin{equation}
 Z_G(Q,Q_{\rm s};{\bf v}) = Q^{|V|/2} \sum_{A \subseteq E} Q^{\ell(A)/2}
 \left( \frac{Q_{\rm s}}{Q} \right)^{\ell_{\rm s}(A)} \prod_{e \in A} x_e
 \label{BLM}
\end{equation}
where we have introduced $x_e = Q^{-1/2} v_e$. In other words, the
weight of a bulk loop (resp.\ a boundary loop) is $n$ (resp.\ $n_{\rm s}$),
subject to the relations
\begin{equation}
 Q = n^2 \qquad
 Q_{\rm s} = n n_{\rm s}
\label{Qn_trans}
\end{equation}

The boundary loop model (\ref{BLM}) was introduced in \cite{JS1}, and
further studied in a more general setting in \cite{JS2}. The emphasis
in Refs.~\cite{JS1,JS2} was on the ferromagnetic case where all $x_e = 1$.
We shall see now that the generalization of these results to the
antiferromagnetic region (with $x_e < 0$) allows to infer the phase
diagram of the boundary chromatic polynomial.

\section{Beraha-Kahane-Weiss theorem}
\label{sec:BKW}

We wish to study the boundary chromatic polynomial in the
thermodynamic limit where $G$ becomes large ($|V| \to \infty$).  In
general, one may take the limit $|V| \to \infty$ through a
recursive family of graphs $G_N$ embedded in the annulus, of width $W$
and circumference $N$, such that $|V| \sim N W$ and $|V_{\rm s}| \sim
N$. In particular one may think of strips of regular lattices
(square, triangular,\ldots), but we emphasize that most of our results
do not depend on the detailed structure of the graph, nor do they
require that it be regular.

In section~\ref{sec:infiniteW} we take the width $W \propto N$, so
that the limiting graph $G_\infty$ is two-dimensional, and the results
\cite{JS1} of conformal field theory (CFT) apply. In
section~\ref{sec:finiteW}, we consider instead $W$ finite, so that
$G_\infty$ is quasi one-dimensional, and we shall see that the main
results hold true in that case as well.

In both cases one may think of the partition function $P_G(Q,Q_{\rm
  s})$ as being built up by a transfer matrix, with time slices
containing $W$ spins. The structure of the transfer matrix has been
discussed in details in \cite{JS1}, and in particular it was shown
that each of its eigenvalues $\lambda_i$ contributes to the partition
function with a non-trivial multiplicity $D_i$ that we shall refer to
as an {\em eigenvalue amplitude}. Hence,
\begin{equation}
 P_G(Q,Q_{\rm s}) = \sum_i D_i (\lambda_i)^N \,.
 \label{eigamp}
\end{equation}
The fact that $D_i \neq 1$ in general can be traced back to the
non-local nature of the loops defining (\ref{BLM}), and to the
periodic boundary conditions in the time direction.

We wish to study the phase diagram of the boundary chromatic
polynomial by locating the accumulation points ${\cal A}$ of the
partition function zeroes $P_G(Q,Q_{\rm s}) = 0$ in the limit $N \to
\infty$.  Following Lee and Yang \cite{LeeYang}, this can be done by fixing
one of the variables $Q$ or $Q_{\rm s}$ (or by imposing a fixed
relation among $Q$ and $Q_{\rm s}$), and letting the remaining variable
(henceforth denoted $z$) take complex values.

Due to the form (\ref{eigamp}) the Beraha-Kahane-Weiss (BKW) theorem
\cite{BKW} applies. Let us call an eigenvalue $\lambda_i$ dominant at $z$
if $|\lambda_i(z)| \ge |\lambda_k(z)|$ for all $k$. The BKW theorem then
states that (under very mild assumptions)
\begin{itemize}
\item $z \in {\cal A}$ is an isolated accumulation point iff there is
  a {\em unique} dominant eigenvalue $\lambda_i$ at $z$ and the
  corresponding amplitude vanishes, $\alpha_i(z)=0$.
 \item $z \in {\cal A}$ forms part of a continuous curve of accumulation points
  iff there are {\em at least two} dominant eigenvalues at $z$. (In other words,
  $z$ is the locus of a level crossing involving a dominant eigenvalue.)
\end{itemize}

It is not in general clear to what extent CFT predictions apply to
complex values of the parameters $Q$ and $Q_{\rm s}$. But at least we
can infer important information about the phase diagram by combining
the BKW theorem \cite{BKW} with the CFT results \cite{JS1} for the
special case of real parameter values.

\section{Phase diagram in the thermodynamic limit}
\label{sec:infiniteW}

It is useful to parametrize the bulk and boundary loop weights as follows
\begin{equation}
 n = 2 \cos(\pi e_0) \,, \qquad
 n_{\rm s} = \frac{\sin((r+1)\pi e_0)}{\sin(r \pi e_0)}
 \label{param_n}
\end{equation}
defining the parameters $e_0$ and $r \in (0,\frac{1}{e_0})$.
The continuum theory then has central charge
\begin{equation}
 c = 1 - \frac{6 e_0^2}{1-e_0}
\label{central}
\end{equation}
The range $e_0 \in [0,\frac12)$ describes the usual
ferromagnetic-paramagnetic transition, corresponding to positive
values of $n$ and $n_{\rm s}$.

We here need the analytic continuation into the range $e_0 \in
(\frac12,1)$, where $n$ and $n_{\rm s}$ become negative. This range
was referred to as the Berker-Kadanoff (BK) phase in \cite{Saleur}.
Inspecting Fig.~\ref{fig:trans} it is easy to see that (\ref{BLM}) is
invariant under a simultaneous sign change of $n$, $n_{\rm s}$, and
$x_e$.  The BK phase therefore corresponds to negative values of
$x_e$, i.e., it describes a part of the antiferromagnetic region of
the Potts model. Its relevance to the chromatic line $v_e = -1$ is due
to the fact that the temperature variable $v_e$ is an {\em irrelevant}
perturbation in the BK phase, in the sense of the renormalization
group. The BK phase therefore controls, for any fixed $Q \in (0,4)$, a
{\em finite} range of values $v_e$. One may therefore expect that at
least for $Q < Q_{\rm c}$, where $Q_{\rm c} \le 4$ is some lattice-dependent
constant, the BK phase will control the chromatic line $v_e = -1$.

To give a little more substance to this general discussion, it is
worthwhile to recall some exact information about the special cases
of the square and triangular lattices. The standard Potts model
($Q_{\rm s}=Q$ and $v_e=v$) is then exactly solvable on the curves
\cite{Baxter_sq,Baxter_tri}
\begin{eqnarray}
 v^2 &=& Q \quad \mbox{(square lattice)} \nonumber \\
 v^3 + 3 v^2 &=& Q \quad \mbox{(triangular lattice)}
\label{critcurves}
\end{eqnarray}
In view of the parametrization (\ref{param_n}) it is more convenient
to rewrite this as
\begin{eqnarray}
 v &=& 2 \cos(\pi e_0) \qquad \qquad \quad \mbox{(square lattice)} \nonumber \\
 v &=& -1 + 2 \cos\left( \frac{2 \pi e_0}{3} \right) \quad
 \mbox{(triangular lattice)}
 \label{critcurves1}
\end{eqnarray}
where $e_0 \in [0,1]$ for the square lattice and $e_0 \in [0,\frac32]$ for
the triangular lattice.

Both analytical and numerical studies of the Potts model with $Q_{\rm
  s} = Q$ and either free or periodic transverse boundary conditions
conclude that the critical exponents along the curves
(\ref{critcurves}) for $e_0 \in [0,1)$ are those predicted by the CFT.
In particular, the central charge is (\ref{central}) as
claimed. Moreover, the exponents for $e_0 \in (\frac12,1)$ are just
the analytic continuations of those valid for the usual ferromagnetic
regime $e_0 \in (0,\frac12)$. This already strongly suggests that the
critical properties for $e_0 \in [0,1)$ are lattice-independent
(universal).%
\footnote{In the case of the triangular lattice, the range $e_0 \in
  (1,\frac32]$ describes a very different CFT \cite{JS_AF} which we
  shall not need further in the present work.}  
This conclusion is further corroborated by the so-called Coulomb gas
approach \cite{Nienhuis} to CFT.

Further studies have established that
for each $Q \in (0,Q_{\rm c})$ the chromatic polynomial indeed
renormalizes to the BK phase, with the following values of $Q_{\rm c}$
for the square \cite{Baxter_AF} and triangular
\cite{Baxter_chrom,Salas_torus} lattices
\begin{eqnarray}
 Q_{\rm c} &=& 3 \qquad \qquad \qquad \quad \quad \,
 \mbox{(square lattice)} \nonumber \\
 Q_{\rm c} &=& 3.8196717312\cdots \quad \ \mbox{(triangular lattice)}
\label{BaxterQc}
\end{eqnarray}

We now return to the main objective of this section, which is to
establish the critical behavior of the boundary chromatic
polynomial. On general grounds, boundary conditions should not modify
bulk RG flows.%
\footnote{This is of course a subtle issue in cases such as this one,
  where the statistical models are not very physical.  In fact, there
  {\em are} some boundary terms that can profoundly affect the
  behavior of flows in the Berker-Kadanoff phase---for instance those breaking the quantum group symmetry in the XXZ chain version of the models. For the boundary
  terms we are considering however---which can be described through
  the boundary Temperley Lieb algebra---no such ``rogue'' behavior
  seems to occur.}
Therefore, we expect the analytic continuation of the CFT results
\cite{JS1} to the range $e_0 \in (\frac12,1)$ to describe the critical
behavior of the boundary chromatic polynomial for $Q \in (0,Q_{\rm
  c})$.

It is convenient to set $e_0=1-\frac{1}{t}$, so that the BK phase
corresponds to $t>2$. The parameter $r$ appearing in (\ref{param_n})
is then constrained to $r \in (0,\frac{t}{t-1})$. We have
\begin{eqnarray}
 n &=& -2 \cos \left( \frac{\pi}{t} \right) \nonumber \\
 n_{\rm s} &=& -\frac{\sin \left( \frac{\big(r(t-1)-1\big) \pi}{t} \right)}
                    {\sin \left( \frac{r(t-1)\pi}{t} \right)}
\label{param_t}
\end{eqnarray}
In this parametrization, $Q=n^2$ [see Eq.~(\ref{Qn_trans})] is nothing
else than the $t$'th Beraha number $B_t$ defined in
(\ref{Berahano}). Real chromatic zeroes have long been known
\cite{Sokal1} to accumulate around $B_t$ for integer values of $t \ge
2$. One major motivation of this work is to show that the special role
played by the Beraha numbers is destroyed by chosing $Q_{\rm s} \neq
Q$.

As explained in \cite{JS1} the detailed transfer matrix structure
implies that each eigenvalue appearing in (\ref{eigamp}) is in fact an
eigenvalue of a modified transfer matrix in which the number of loops
winding around the periodic direction of the annulus (i.e., which
are non-homotopic to a point) is fixed. Each eigenvalue can thus be
labelled by the corresponding number of winding loops
$L=0,2,4,\ldots$, as $\lambda_i^{(L)}$. By the definition of the Potts
model and the medial graph ${\cal M}$, the corresponding number of
winding clusters is $L/2$. In each sector with $L>0$, the dominant
eigenvalue corresponds to the outermost of the winding loops being
constrained to be a boundary loop (i.e., we can restrict to what was
called the ``blobbed sector'' in \cite{JS1,JS2}).

The existence of $L$ winding loops corresponds in CFT to the insertion
of a pair of so-called $L$-leg operators ${\cal O}_L$ at the
extremities of the strip; the extremities are subsequently glued
together to form the annulus with periodic boundary conditions in the
time direction. The asymptotic scaling for $W \gg 1$ of the dominant
eigenvalue $\lambda_0^{(L)}$ in each sector $L$ is then fixed by CFT
as \cite{Cardy84}
\begin{equation}
 \frac{\lambda_0^{(L)}}{\lambda_0^{(0)}} =
 \exp \left( -\frac{\pi h_L}{W} \right) + \ldots
 \label{fss}
\end{equation}
where the dots on the right-hand side represent terms that decay faster
than $\exp(-{\rm const}/W)$.

The constant $h_L$ appearing in (\ref{fss}) is the so-called conformal
weight of the $L$-leg operator (in the ``blobbed sector'') whose value
has been established in \cite{JS1,Kostov}. After the analytic continuation
implied by the parametrization (\ref{param_t}), this reads
\begin{equation}
 h_L = \frac{1}{4t} \left( L^2 - 2 r L (t-1) + (r^2-1)(t-1)^2 \right)
 \label{domexp}
\end{equation}
The corresponding eigenvalue amplitude has been derived rigorously in
\cite{JS1}:
\begin{equation}
 D_L = \left \lbrace \begin{array}{ll}
 1 & \mbox{for } L=0 \\
 n_{\rm s} U_{L-1}(n/2) - U_{L-2}(n/2) & \mbox{for } L > 0 \\
 \end{array} \right.
 \label{domeigamp}
\end{equation}
where $U_j(x)$ is the $j$'th order Chebyshev polynomial of the second kind.

Note that in the probabilistic regime ($e_0 \in [0,\frac12]$) the
continuum limit is dominated by $L=0$.  This is no longer true in the
BK phase, where for any $t$ there is at least one of the exponents
$h_L$ taking negative values. The most negative exponent determines
the most ``probable'' number of winding loops. This situation is
clearly counter-intuitive from a probabilistic point of view, but it
is made possible by the appearance of negative Boltzmann weights. Note
also that the invariance of (\ref{BLM}) under a simultaneous sign
change of $n$, $n_{\rm s}$, and $x_e$ is not sufficient to make all
weights positive.

Using (\ref{fss}), dominant level crossings of transfer matrix
eigenvalues correspond asymptotically (for $W \gg 1$) to level
crossings of the conformal weights $h_L$. We can thus read
directly from (\ref{domexp}) the necessary and sufficient criterion
for the second part of the BKW theorem. Indeed,
level crossings involving the dominant $L$-leg sector occur when
\begin{equation}
 h_L = h_{L+2} \Leftrightarrow r = \frac{L+1}{t-1}
\label{h_criterion}
\end{equation}
with $L \le t-1$. In particular, $h_L$ is the most negative exponent
for $r \in (\frac{L-1}{t-1},\frac{L+1}{t-1})$.

Similarly, the necessary and sufficient criterion for the first part
of the BKW theorem is read off from (\ref{domeigamp}). Indeed, the
amplitude of the dominant $L$-leg sector vanishes when
\begin{equation}
 D_L = 0 \Leftrightarrow r = \frac{L}{t-1}
\label{D_criterion}
\end{equation}
with $L=2,4,6,\ldots$.

\begin{figure}
\begin{center}
\includegraphics[width=7cm,angle=270]{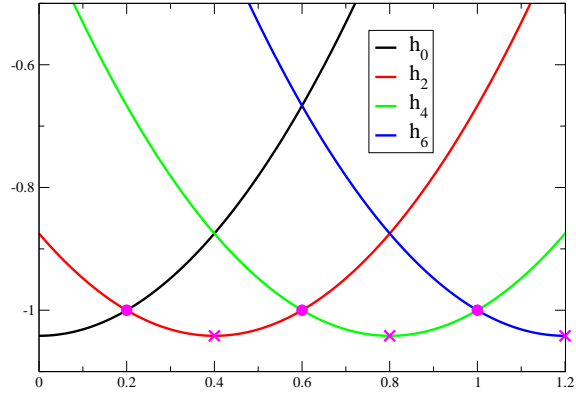}
\caption{Conformal weights $h_L$ as functions of the parameter $r$ for
the case $t=6$. Dominant level crossings and vanishing dominant amplitudes
are shown respectively as solid circles and crosses.}
\label{fig:levels}
\end{center}
\end{figure}

These phenomena are illustrated in Fig.~\ref{fig:levels} for the case
$t=6$ (the $Q=3$ state Potts model).

For any fixed $n$, phase transitions will therefore take place
for $r=s/(t-1)$ and integer $s \in (0,t]$. The corresponding
value of the boundary parameter is
\begin{equation}
 n_{\rm s} = - \frac{\sin \left( \frac{(s-1)\pi}{t} \right)}
                   {\sin \left( \frac{s \pi}{t} \right)}
\end{equation}
For even $s$ this corresponds to a vanishing dominant amplitude, and
for odd $s$ to a dominant level crossing. The corresponding value of
the dominant exponent (\ref{domexp}) is $h_L = -\frac{(t-1)^2}{4t}$
for any even $s$, and $h_{\rm L} = \frac{2-t}{4}$ for any odd $s$.

The $N\to\infty$ limiting curve of accumulation points of partition
function zeroes in the complex $Q_{\rm s}$ plane (in the vicinity of
the real $Q_{\rm s}$ axis) can now be inferred from the BKW theorem:
For even $s$ one has an isolated real accumulation point, and for odd
$s$ a continuous curve of accumulation points intersects the real
axis.

In the example $t=6$ of Fig.~\ref{fig:levels}, the
transitions at $r=\frac15,\frac25,\frac35,\frac45,1,\frac65$
correspond to the following numbers of boundary colors: $Q_{\rm s} =
0,1,\frac32,2,3,\infty$.

The discussion following (\ref{D_criterion}) has subsumed that we are
interested in the phase diagram for fixed $Q$ and varying $Q_{\rm s}$.
But of course the criteria (\ref{h_criterion})--(\ref{D_criterion})
for phase transitions hold true for other situations as well. In
particular, the following few useful cases correspond to simple relations
between $r$ and $t$:
\begin{equation}
 \begin{tabular}{lll}
 $Q_{\rm s}=Q$   & : & $r = 1$ \\
 $Q_{\rm s}=Q-1$ & : & $r = (t-2)/(t-1)$ \\
 $Q_{\rm s}=Q-2$ & : & $r = (t-2)/(2t-2)$ \\
 $Q_{\rm s}=Q-\sqrt{Q}$ & : & $r = 1/2$ \\
 $Q_{\rm s}=0$   & : & $r = 1/(t-1)$ \\
 $Q_{\rm s}=1$   & : & $r = 2/(t-1)$ \\
 $Q_{\rm s}=2$   & : & $r = (t+2)/(2t-2)$ \\
 $Q_{\rm s}=\frac12 Q$ & : & $r = t/(2t-2)$ \\
 \end{tabular}
\end{equation}
For all of these, (\ref{h_criterion})--(\ref{D_criterion}) yield
phase transitions located at {\em integer} values of $t$ (i.e., at
the Beraha numbers $B_t$), but this needs of course not be the case
for more general choices of $Q_{\rm s}$.

\section{Quasi one-dimensional case}
\label{sec:finiteW}

We now turn to the quasi one-dimensional geometry where the circumference
of the annulus $N \to \infty$, while is width $W$ is kept fixed and finite.
In that case, the possible number of winding loops is constrained by
$L \le 2W$.

Eq.~(\ref{domeigamp}) for the eigenvalue amplitudes was in fact
derived combinatorially for finite $W$, and so remains valid in this
case.  On the other hand, Eq.~(\ref{domexp}) must be discarded, since
its derivation supposed the validity of conformal field theory.
However, the pleasant surprise is that even for finite $W$ the
dominant eigenvalues in the $L$ and $(L+2)$ leg sectors cross
exactly for the values of $r$ and $t$ given by (\ref{h_criterion}).

This coincidence follows from representation theory of the underlying
boundary Temperley-Lieb algebra. While this algebra is semi-simple for
generic values of the parameters, it admits families of degeneracy
points where generically irreducible representations merge into larger
indecomposable representations. Results in \cite{MSblob} guarantee
that this occurs for finite values of $W$ exactly at the same values
that lead to the coincidences (\ref{h_criterion}) of the conformal
weights in the continuum limit.
 
When $r=1$ ---in the original parametrization (\ref{param_n})---this
can be understood somewhat more easily by using quantum group
representation theory \cite{PS_quant}, as the generic $U_q(sl(2))$
representations for sectors $L$ and $L+2$, of spin $j=L/2$ and $j=L/2+1$,
merge into larger indecomposable representations. When $r$ is integer
larger than one, this can be explained similarly by the construction
of section 5 in Ref.~\cite{JS1}. Indeed, there the effect of the
boundary weight $n_{\rm s}$ was obtained algebraically by adding $r$
extra strands on the outside of the annulus, subject to the action of
a certain symmetrizer. Thus, the boundary loop model (\ref{BLM}) with
$r$ integer is a special case of the standard loop model in which only
the weight $n$ appears. The latter is known to have an $U_q(sl(2))$
quantum group symmetry \cite{PS_quant}, and this in fact implies that
(\ref{h_criterion}) still holds true. The presence of exact
coincidences at arbitrary $r$ can maybe be interpreted in terms of
some quantum group---the commutant of the boundary Temperley-Lieb
algebra---but we will not discuss this here.

The key results of section~\ref{sec:infiniteW} therefore remain  
valid, up to two subtle effects to be discussed below.

\begin{figure}
\begin{center}
\includegraphics[width=7cm,angle=270]{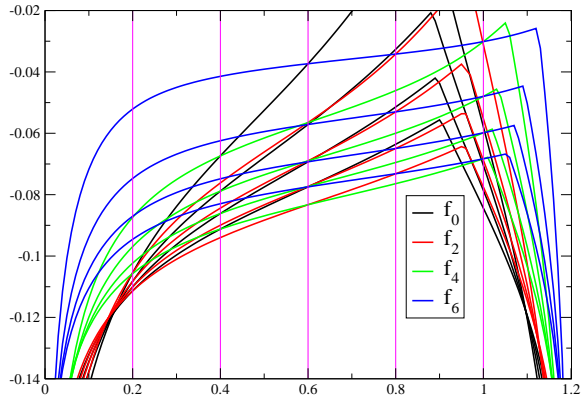}
\caption{Leading free energies $f_L$ in the sectors $L=0,2,4,6$ as functions
of $r \in (0,\frac65)$. The boundary loop model is here defined on the square
lattice, along the BK critical curve, and the parameter $t=6$. Four different
system sizes ($W=8,10,12,14$) are shown, the largest size corresponding to
the lowermost curves. The vertical lines are guides to the eye.}
\label{fig:free}
\end{center}
\end{figure}

To make this conclusion more accessible to readers unacquainted with
quantum groups we turn to a
numerical verification. Fig.~\ref{fig:free} shows the leading free energies
$f_L = -\frac{1}{W} \log \lambda_0^{(L)}$ in the $L$-leg sector, as functions
of $r$ in the parametrization (\ref{param_t}), for four different values
of $W$. The results were obtained for the square lattice in the diagonal
geometry defined in \cite{JS1}, along the curve (\ref{critcurves1}) with
$e_0 \in (\frac12,1)$, i.e., within the BK phase. Results for other lattices
would be similar, provided that one remains inside the domain of attraction of
the BK phase.

For each $W$, the dominant level crossings are seen to occur exactly
as predicted by (\ref{h_criterion}). More generally, the $r$ values
singled out by (\ref{h_criterion})--(\ref{D_criterion}) are seen to
be the loci of subdominant level crossings as well, as would be expected
from an underlying quantum group symmetry.

Fig.~\ref{fig:free} was made for the choice $t=6$ (the $Q=3$ state
Potts model), so that it is the precise finite-size analogue of
Fig.~\ref{fig:levels}. Other, non-integer choices of $t$ were found to
lead to the same conclusions.

We still need to discuss the two subtle effects referred to above. The
first one is that if the annulus is too narrow ($2W < \lfloor t \rfloor$)
to accomodate the number of legs required by dominant sector with the
largest $L$ predicted by (\ref{h_criterion}), the corresponding level
crossings will simply be absent, and the $2W$-leg sector will remain
dominant for the corresponding values of the parameter $r$.

The second effect is that Fig.~\ref{fig:free} gives clear evidence that
when $r$ becomes too large, there is an internal level crossing in each
$L$-leg sector, visible as a cusp in the curves. To the right of these
cusps the pattern of dominance may change. A detailed analysis of
the loci of the cusps reveals that their position tends to $r=1$ as
$W \to \infty$, independently of the value of $L$. Moreover, for
$r \in (1,\frac{t}{t-1})$ it is the $L=0$ sector that will be dominant
for large enough $W$.

\section{Numerical verifications}
\label{sec:numerics}

To conclude this paper, we wish to check that the predictions of
sections~\ref{sec:infiniteW}--\ref{sec:finiteW} agree with existing
numerical results on the limiting curves ${\cal A}$ of chromatic
zeroes. The goal of this comparison is furthermore to convince the
reader that our results are:
\begin{enumerate}
 \item Lattice independent;
 \item Independent of $v_e$, as long as we are in the BK phase;
 \item Correct for various choices of $Q_{\rm s}$.
\end{enumerate}

\begin{figure}
\begin{center}
\includegraphics[width=10cm,angle=0]{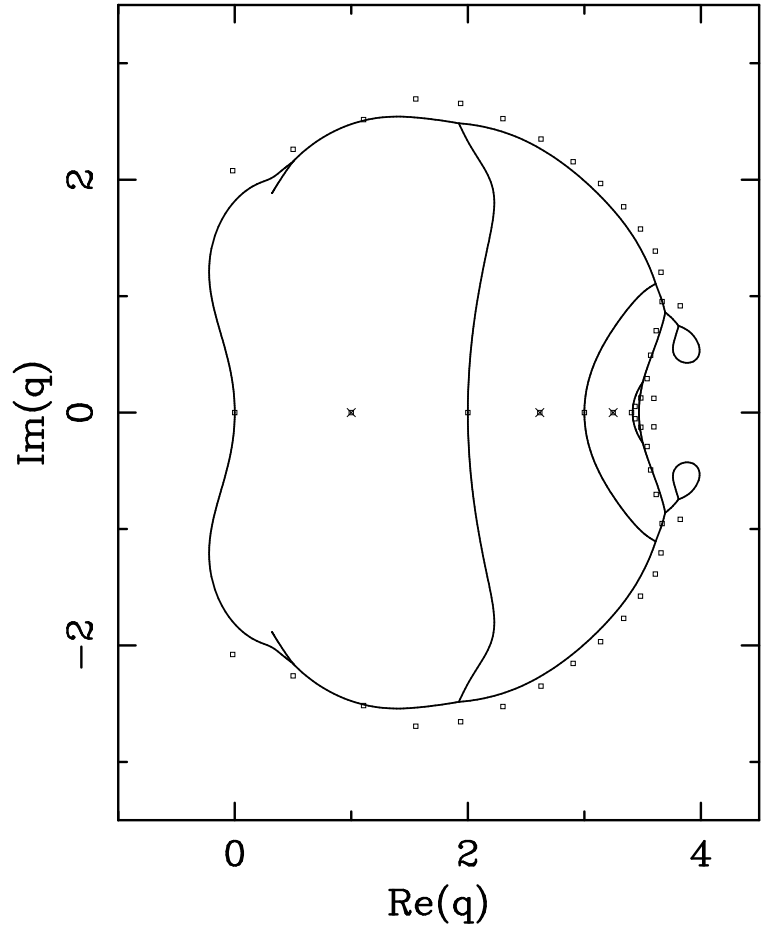}
\caption{Zeroes in the complex $Q$ plane of the triangular-lattice
  chromatic polynomial on an $W \times N$ annulus for $W=7$ and
  $N=35$, and their accumulation points as $N \to \infty$. The
  boundary parameter $Q_{\rm s} = Q$. Taken from Figure 7 of
  Ref.~\cite{Salas_ring}.}
\label{fig:tri7R}
\end{center}
\end{figure}

Fig.~\ref{fig:tri7R} shows the accumulation points ${\cal A}$ for the
triangular-lattice chromatic polynomial on an annulus of width $W=7$.
Transverse boundary conditions are free, so that $Q_{\rm s}=Q$. The agreement
with the predictions (\ref{h_criterion})--(\ref{D_criterion}) for the real
accumulation points is perfect. There is one additional real accumulation
point at $Q_{\rm c}(W) = 3.4682618071\cdots$ which is a finite-size analogue
of $Q_{\rm c}$ discussed in section~\ref{sec:infiniteW}. As $W \to \infty$
we expect $Q_{\rm c}(W) \to Q_{\rm c}$ given by (\ref{BaxterQc}).

\begin{figure}
\begin{center}
\includegraphics[width=9cm,angle=0]{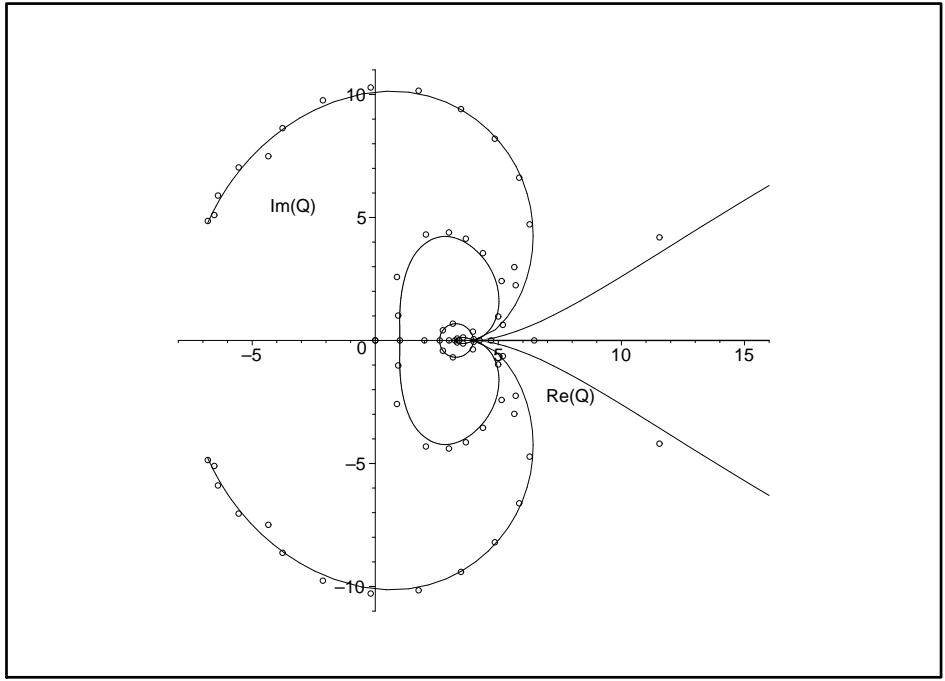}
\caption{Zeroes in the complex $Q$ plane of a square-lattice Potts
  model along the curve (\ref{critcurves}) on an $W \times N$ annulus
  for $W=3$ and $N=26$, and their accumulation points as $N \to
  \infty$. The boundary parameter $Q_{\rm s} = Q-1$. Taken from Figure
  5 of Ref.~\cite{Chang}.}
\label{fig:sqsd3R}
\end{center}
\end{figure}

Fig.~\ref{fig:sqsd3R} shows the accumulation points ${\cal A}$ of
partition function zeroes for a square-lattice Potts model along the
curve (\ref{critcurves}). The geometry is that of an annulus of width
$W=3$ with free transverse boundary conditions. However, all vertices
on the outer rim of the annulus are connected to an extra exterior
vertex. Therefore, the vertices on the outer rim (call them $V_{\rm s}$)
support spins which are effectively constrained to take only
$Q_{\rm s}=Q-1$ different values (since they must be different from the
value of the exterior spin). The partition function on the graph just
described is therefore equal to $Q Z_G(Q,Q_{\rm s}=Q-1;v_e = \pm \sqrt{Q})$
in our notation, where now $G$ is just an ordinary annulus of width $W$,
with no extra exterior vertex.

Once again, the agreement with the predictions
(\ref{h_criterion})--(\ref{D_criterion}) for the real accumulation
points is perfect.  In particular, it follows easily from the predictions that 
the loci of isolated real accumulation points
and curves of accumulation points intersecting the real $Q$-axis are
 swapped between Figs.~\ref{fig:tri7R} and \ref{fig:sqsd3R}.
Along the curve (\ref{critcurves}) we would expect
the BK phase to terminate only at $Q_{\rm c}=4$. Thus, the phase
transition corresponding to the largest possible $L$-sector becoming
dominant is limited by the available width as $L \le 2W$. This is
again in perfect agreement with Fig.~\ref{fig:sqsd3R}. Similar
agreements are found with the numerical results for real accumulation
points given in \cite{Chang} in the case $W=4$ (for which the complete
limiting curve ${\cal A}$ was not computed).

\begin{figure}
\begin{center}
\includegraphics[width=12cm,angle=270]{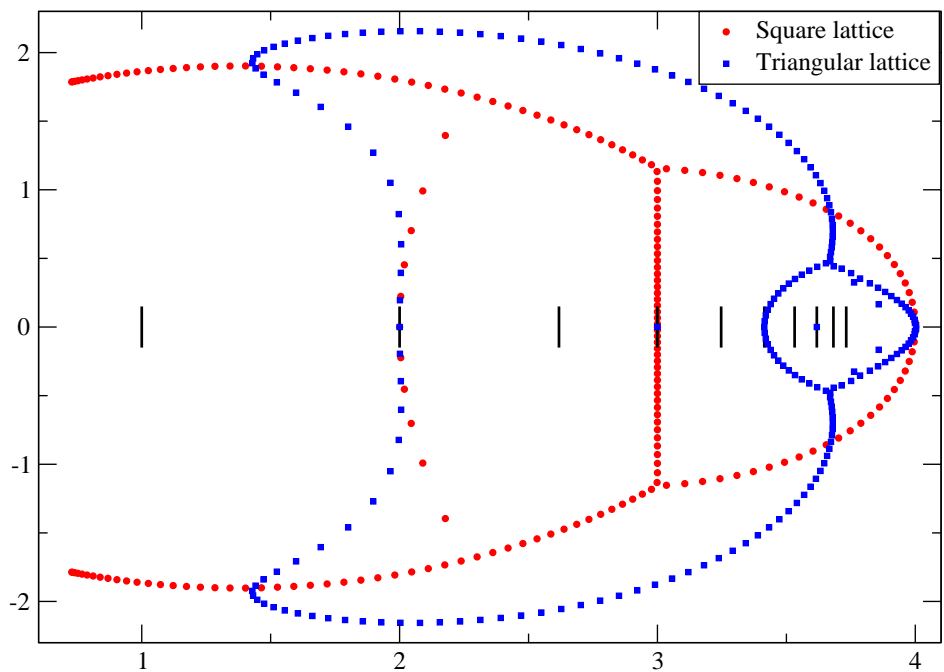}
\caption{Zeroes in the complex $Q$ plane of the $Q_{\rm s}=Q-2$
  boundary chromatic polynomials on an $W \times N$ annulus, with $W=2$
  and $N=100$, for both the square and the triangular lattice.  The
  black vertical lines indicate the positions of the Beraha numbers
  (\ref{Berahano}).}
\label{fig:sqtri2R}
\end{center}
\end{figure}

As a final check, we have computed the boundary chromatic polynomials
with $Q_{\rm s} = Q-2$ on an $W \times N$ annulus for $W=2$ and
$N=100$, for both the square and the triangular lattice. Their zeroes
in the complex $Q$ plane are shown in Fig.~\ref{fig:sqtri2R}.  The
agreement between Eqs.~(\ref{h_criterion})--(\ref{D_criterion}) and
the real accumulation points for the triangular lattice is striking.
Notice in particular that we predict in general that only Beraha
numbers of even order, viz.~$B_t$ with $t=4,6,8,10,\ldots$, can appear as
accumulation points on the real $Q$ axis.  For the square lattice, the
branch cutting the real axis at $Q=3$ marks the termination of the BK
phase, in agreement with (\ref{BaxterQc}); to the right of this branch
one does not observe any further structure as expected.

\section{Conclusion}

To summarize, we have introduced a new graph coloring problem---the
boundary chromatic polynomial---and identified the loci of phase
transitions for real values of the parameters $Q$ and $Q_{\rm s}$.
Our results are lattice independent, and valid not only on the
chromatic line but in the entire Berker-Kadanoff phase.

While we have provided a number of striking numerical tests that
validate our analytical predictions, we believe we have left ample
space for further numerical investigations of the boundary chromatic
zeroes for families of graphs embedded in the annulus.

A straightforward extension of the work presented here would be to
consider graphs on an annulus for which bulk spins can take values
$1,2,\ldots,Q$, whereas spins on the outer (resp.\ inner) rim of the
annulus are constrained to take values $1,2,\ldots,Q_{\rm o}$ (resp.\
$1,2,\ldots,Q_{\rm i}$). Note that in the cluster expansion analogous
to (\ref{FK}), the number of spin values accessible to clusters
touching both rims can be taken as a further independent variable
$Q_{\rm b}$, not necessarily equal to ${\rm min}(Q_{\rm o},Q_{\rm
  i})$.

Recent work on the corresponding two-boundary loop model furnishes the
results for the eigenvalue amplitudes \cite{JS2} and the critical
exponents \cite{DJS}, analogous to (\ref{domexp})--(\ref{domeigamp})
of this article.  The phase diagram for real parameter values $Q$,
$Q_{\rm o}$, $Q_{\rm i}$, $Q_{\rm b}$ can therefore be worked out
along the lines presented here.

\subsubsection*{Acknowledgments}

We thank the authors of Refs.~\cite{Salas_ring,Chang} for the
permission to reproduce Figs.~\ref{fig:tri7R}--\ref{fig:sqsd3R}. This
work was supported by the European Community Network ENRAGE (grant
MRTN-CT-2004-005616), by the Agence Nationale de la Recherche
(grant ANR-06-BLAN-0124-03), and by the European Science Foundation
network program INSTANS. One of us (JLJ) further thanks the
Isaac Newton Institute for Mathematical Sciences, where part of this
work was done, for hospitality.

\end{document}